\documentstyle[twoside]{article}

\catcode`\@=11
\long\def\@makefntext#1{
\protect\noindent \hbox to 3.2pt {\hskip-.9pt  
$^{{\eightrm\@thefnmark}}$\hfil}#1\hfill}		

\def\@makefnmark{\hbox to 0pt{$^{\@thefnmark}$\hss}}	
	
\def\ps@myheadings{\let\@mkboth\@gobbletwo
\def\@oddhead{\hbox{}
\rightmark\hfil\eightrm\thepage}   
\def\@oddfoot{}\def\@evenhead{\eightrm\thepage\hfil
\leftmark\hbox{}}\def\@evenfoot{}
\def\sectionmark##1{}\def\subsectionmark##1{}}

\oddsidemargin=\evensidemargin
\newcounter{sectionc}\newcounter{subsectionc}\newcounter{subsubsectionc}
\renewcommand{\section}[1] {\vspace{12pt}\addtocounter{sectionc}{1} 
\setcounter{subsectionc}{0}\setcounter{subsubsectionc}{0}\noindent 
	{\tenbf\thesectionc. #1}\par\vspace{5pt}}
\renewcommand{\subsection}[1] {\vspace{12pt}\addtocounter{subsectionc}{1} 
	\setcounter{subsubsectionc}{0}\noindent 
	{\bf\thesectionc.\thesubsectionc. {\kern1pt \bfit #1}}\par\vspace{5pt}}
\renewcommand{\subsubsection}[1] {\vspace{12pt}\addtocounter{subsubsectionc}{1}
	\noindent{\tenrm\thesectionc.\thesubsectionc.\thesubsubsectionc.
	{\kern1pt \tenit #1}}\par\vspace{5pt}}

\newcounter{appendixc}
\newcounter{subappendixc}[appendixc]
\newcounter{subsubappendixc}[subappendixc]
\renewcommand{\thesubappendixc}{\Alph{appendixc}.\arabic{subappendixc}}
\renewcommand{\thesubsubappendixc}
	{\Alph{appendixc}.\arabic{subappendixc}.\arabic{subsubappendixc}}

\renewcommand{\appendix}[1] {\vspace{12pt}
        \refstepcounter{appendixc}
        \setcounter{figure}{0}
        \setcounter{table}{0}
        \setcounter{lemma}{0}
        \setcounter{theorem}{0}
        \setcounter{corollary}{0}
        \setcounter{definition}{0}
        \setcounter{equation}{0}
        \renewcommand{\thefigure}{\Alph{appendixc}.\arabic{figure}}
        \renewcommand{\thetable}{\Alph{appendixc}.\arabic{table}}
        \renewcommand{\theappendixc}{\Alph{appendixc}}
        \renewcommand{\thelemma}{\Alph{appendixc}.\arabic{lemma}}
        \renewcommand{\thetheorem}{\Alph{appendixc}.\arabic{theorem}}
        \renewcommand{\thedefinition}{\Alph{appendixc}.\arabic{definition}}
        \renewcommand{\thecorollary}{\Alph{appendixc}.\arabic{corollary}}
        \renewcommand{\theequation}{\Alph{appendixc}.\arabic{equation}}
        \noindent{\tenbf Appendix \theappendixc #1}\par\vspace{5pt}}
\newcommand{\subappendix}[1] {\vspace{12pt}
        \refstepcounter{subappendixc}
        \noindent{\bf Appendix \thesubappendixc. {\kern1pt \bfit #1}}
	\par\vspace{5pt}}
\newcommand{\subsubappendix}[1] {\vspace{12pt}
        \refstepcounter{subsubappendixc}
        \noindent{\rm Appendix \thesubsubappendixc. {\kern1pt \tenit #1}}
	\par\vspace{5pt}}

\newcommand{\textlineskip}{\baselineskip=13pt}
\newcommand{\smalllineskip}{\baselineskip=10pt}

\def\eightcirc{
\begin{picture}(0,0)
\put(4.4,1.8){\circle{6.5}}
\end{picture}}
\def\eightcopyright{\eightcirc\kern2.7pt\hbox{\eightrm c}} 

\newcommand{\copyrightheading}[1]
	{\vspace*{-2.5cm}\smalllineskip{\flushleft
	{\footnotesize International Journal of Modern Physics E, #1}\\
	{\footnotesize $\eightcopyright$\, World Scientific Publishing
	 Company}\\
	 }}

\newcommand{\publisher}[2]{{\begin{center}\footnotesize\smalllineskip 
	Received #1\\
	Revised #2
	\end{center}
	}}

\def\abstracts#1#2#3{{
	\centering{\begin{minipage}{4.5in}\baselineskip=10pt\footnotesize
	\parindent=0pt #1\par 
	\parindent=15pt #2\par
	\parindent=15pt #3
	\end{minipage}}\par}} 

\renewenvironment{thebibliography}[1]
	{\frenchspacing
	 \ninerm\baselineskip=11pt
	 \begin{list}{\arabic{enumi}.}
        {\usecounter{enumi}\setlength{\parsep}{0pt}     
	 \setlength{\leftmargin 12.7pt}{\rightmargin 0pt} 
         \setlength{\itemsep}{0pt} \settowidth
	{\labelwidth}{#1.}\sloppy}}{\end{list}}

\newcounter{itemlistc}
\newcounter{romanlistc}
\newcounter{alphlistc}
\newcounter{arabiclistc}

\newcommand{\fcaption}[1]{
        \refstepcounter{figure}
        \setbox\@tempboxa = \hbox{\footnotesize Fig.~\thefigure. #1}
        \ifdim \wd\@tempboxa > 5in
           {\begin{center}
        \parbox{5in}{\footnotesize\smalllineskip Fig.~\thefigure. #1}
            \end{center}}
        \else
             {\begin{center}
             {\footnotesize Fig.~\thefigure. #1}
              \end{center}}
        \fi}

\newcommand{\tcaption}[1]{
        \refstepcounter{table}
        \setbox\@tempboxa = \hbox{\footnotesize Table~\thetable. #1}
        \ifdim \wd\@tempboxa > 5in
           {\begin{center}
        \parbox{5in}{\footnotesize\smalllineskip Table~\thetable. #1}
            \end{center}}
        \else
             {\begin{center}
             {\footnotesize Table~\thetable. #1}
              \end{center}}
        \fi}

\def\@citex[#1]#2{\if@filesw\immediate\write\@auxout
	{\string\citation{#2}}\fi
\def\@citea{}\@cite{\@for\@citeb:=#2\do
	{\@citea\def\@citea{,}\@ifundefined
	{b@\@citeb}{{\bf ?}\@warning
	{Citation `\@citeb' on page \thepage \space undefined}}
	{\csname b@\@citeb\endcsname}}}{#1}}

\newif\if@cghi
\def\cite{\@cghitrue\@ifnextchar [{\@tempswatrue
	\@citex}{\@tempswafalse\@citex[]}}
\def\citelow{\@cghifalse\@ifnextchar [{\@tempswatrue
	\@citex}{\@tempswafalse\@citex[]}}
\def\@cite#1#2{{$\null^{#1}$\if@tempswa\typeout
	{IJCGA warning: optional citation argument 
	ignored: `#2'} \fi}}

\def\pmb#1{\setbox0=\hbox{#1}
	\kern-.025em\copy0\kern-\wd0
	\kern.05em\copy0\kern-\wd0
	\kern-.025em\raise.0433em\box0}

\def\fnt#1#2{\footnotetext{\kern-.3em
	{$^{\mbox{\scriptsize #1}}$}{#2}}}

\def\fpage#1{\begingroup
\voffset=.3in
\thispagestyle{empty}\begin{table}[b]\centerline{\footnotesize #1}
	\end{table}\endgroup}

\def\runninghead#1#2{\pagestyle{myheadings}
\markboth{{\protect\footnotesize\it{\quad #1}}\hfill}
{\hfill{\protect\footnotesize\it{#2\quad}}}}
\font\tenrm=cmr10
\font\tenit=cmti10 
\font\tenbf=cmbx10
\font\bfit=cmbxti10 at 10pt
\font\ninerm=cmr9

\font\eightrm=cmr8

\def\qed{\hbox{${\vcenter{\vbox{			
   \hrule height 0.4pt\hbox{\vrule width 0.4pt height 6pt
   \kern5pt\vrule width 0.4pt}\hrule height 0.4pt}}}$}}

\def\bsc{{\sc a\kern-6.4pt\sc a\kern-6.4pt\sc a}}	
\def\bflatex{\bf L\kern-.30em\raise.3ex\hbox{\bsc}\kern-.14em 
T\kern-.1667em\lower.7ex\hbox{E}\kern-.125em X} 

\begin{document}

\runninghead{Nonlinear Effects in Nuclear Cluster Problem} 
{Nonlinear Effects in Nuclear Cluster Problem}

\normalsize\textlineskip
\thispagestyle{empty}
\setcounter{page}{1}

\copyrightheading{}			

\vspace*{0.88truein}

\fpage{1}
\centerline{\bf NONLINEAR EFFECTS IN NUCLEAR CLUSTER PROBLEM}

\vspace*{0.37truein}
\centerline{\footnotesize   V.G.~KARTAVENKO\footnote{Permanent address: 
Bogoliubov Theoretical Laboratory, Joint
Institute for Nuclear Reasearch, Dubna, Moscow Region, 141980, Russia;
E-mail: kart@thsun1.jinr.ru},
K.A.~GRIDNEV\footnote{Permanent address: Institute of Physics, St.
Petersburg State University, Saint Petersburg, 198904, Russia;
E-mail: gridnev@phim.niif.spb.su}, and W.~GREINER}
\vspace*{0.015truein}
\centerline{\footnotesize\it 
Intstitut f{\"u}r Theoretische Physik,
Johann Wolfgang G{\"o}the Universit{\"a}}
\baselineskip=10pt
\centerline{\footnotesize\it Frankfurt am Main, D-60054, Germany}
\vspace*{0.225truein}
\publisher{16 December 1997}

\vspace*{0.21truein}
\abstracts{Some nonlinear 
aspects of a cluster phenomenon in nuclei 
are considered using of cubic Nonlinear Schr\"odinger Equation and
Korteveg de Vries Equation.
We discuss the following possible nonlinear effects:
i) the decribing clusters as solitons;
ii) an anomalous large angle scattering of $\alpha$-particles by light and
intermediate nuclei; 
iii) stable vortical objects;
iv) and dynamical clusterization in the presence of instability }{}{}

\vspace*{1pt}\textlineskip      
\section{Motivation}         
\vspace*{-0.5pt}
\noindent
Clusters is a very general phenomenon. They exist not only in nuclei and
atoms, but also in subnuclear and macro physics (from small drops to
atmosphere, stars and galaxes)
(see matherials and the summary talk of Prof.~W.~Greiner
at CLUSTERS'93 Conference 
(Santorini,Greece)\cite{Santorini93}
 
There exist very different theoretical methods, developed in these fields.
However there are only few basic physical ideas, and most of methods
makes deal with nonlinear partial differential equations.

The  aim  of  the  present paper  is  to show that very different
effects of nuclear cluster physics could be explained by
using of cubic Nonlinear Schr\"odinger Equation and
Korteveg de Vries Equation.    

The papers is organized as follows.

In Sec.2 we consider general aspects of nonlinear approach for nuclei.
An anomalous large angle scattering of $\alpha$-particles by light and
intermediate nuclei are discussed.             

A way to descrive clusters as stable nonlinear localized 
waves (solitons) is discussed in Sec.3.
Section 4 is devoted to stable vortical exitations.
Dynamical clusterisation 
and break-up of nuclear system are considered from the point
of view of nonlinear dynamics in Sec.5
Last section is a short summary.   

\eject

\vspace*{1pt}\textlineskip	
\section{Nonlinear Approach for Nuclei}		
\vspace*{-0.5pt}
\noindent
The general philosophy giving birth to the NoSE is a rather simple one.
Previous studies\cite{r1} into the anomalous large angle scattering (ALAS) of
$\alpha$-particles by light and intermediate ($\alpha$-cluster)
nuclei that the rise in $d\sigma/d\Omega$ (elastic) for backward
angles~--~together with the measured resonances, $E_x$, and decay
widths, $\Gamma_x$, of the dinuclear ($\alpha$-cluster) system~--~may
be described by (the same parameter of) a crude effective surface
potential (ESP) arrived at by supplementing the optical model
interaction, $V_{OM}$, by a repulsive (hard or soft) core, $V_r$:
$$ V_{ESP} = V_{OM} + V_r $$
$V_{OM}$ facilitates the correct evaluation of $d\sigma/d\Omega$ for
forward angles while $V_r$ gives the additional potential contribution
allowing for an appropriate description of the measured $E_x$, $\Gamma_x$
and of the backward rise in $d\sigma/d\Omega$ as observed for
$\theta_{cm} > 90^\circ$.

Similar quasi--molecular features as in these $\alpha$-scattering
experiments have also been measured in elastic collisions between
heavy ions. Knowing that semi--classical and in particular hydrodynamical
approaches work nicely when applied to heavy ion physics, it appears
very suggestive to try to model the elastic scattering of two heavy
ions in terms of colliding liquid drops with diffuse surfaces.

Assuming that the ``membranes``/surfaces of the two drops remain
impenetrable, one thus obtain automatically a compression of the
densities in the surface regions of the two touching drops. The
physically motivated boundary condition is then that this
compession~---~giving rise to a repulsive spring--force proportional
to $V_r$~---~vanishes for separations
$$ r > r_0 = R_1 + R_2 + (a_1 + a_2) / 2 $$
between the two ions/drops, i.e. $V_r \stackrel{r > r_0}{\rightarrow} 0$.
$R_i$ and $a_i$ denote radius and diffuseness of the i-th ion; $i=1,2$.
Thus the dynamical interaction between the two nuclei is for $r > r_0$
characterized by their unperturbed/uncompressed asymptotic densities
$\rho_{0i}(R_i,a_i)$~---~corresponding to the measured (charge)
densities~---~and for $r < r_0$ by the perturbed/compressed densities
$\rho_i(r;R_i,a_i)$.

The central equation emerging from such considerations together with
the use of the Euler equation and some algebra in liaison with the
neglect of higher order derivatives is the nonlinear Schroedinger
equation (NoSE)\cite{r1}
$$ -\frac{\hbar^2}{2m}\nabla^2\Psi+\nabla\Psi-
   C(1-\frac{\rho}{\rho_0})\Psi = E\Psi, \hspace{8mm}
  \rho=\mid\Psi\mid^2, \hspace{8mm}
  \rho_0=\rho_{01}+\rho_{02}  \eqno(1)
$$
where we use the conventional notation with $C=K/9$ for the nuclear
compressibility, $\rho$ and $\rho_0$ denote the perturbated and
unperturbated densities respectively. (Usually $K$ is taken to apply
to the bulk of nuclear matter while we consider only compressions
in the surface regions.)

It  turned out that Eq. (1) is a particular solution  of  a  general
equation suggested by Bialynsky-Birula and  Mysielsky\cite{r3}.  Using
instead of cubic term $\epsilon \log (a\left|\matrix{\psi }\right|^{2})\psi $
they formulated  the  nonlinear
quantum mechanics without any contradiction in  classical  quantum
mechanics. By the way, Gaehler, Klein and Zeilinger\cite{r4}  attempted
to find the constant $\epsilon$ according to the predictions  made  in  the
work\cite{r3}. But it  was  made  without  any  success.  Our  approach
supposes the constant $C=K/9$ or $\epsilon$ ( here  both  constants  are  the
same) to be the compressibility constant. It  was  confirmed  from
the  values  of  the  constant $K$,   obtained   from    different
experiments with the help of this approach $(K \approx 270 MeV)$ and  also
by the recent work of J.Braecher\cite{r5}.  J.Braecher  considered  the
nonlinear term  in  the  NoSE  from  the  point  of  view  of  the
information theory. According to the work of  Shannon  and  Weaver
\cite{r6} the information $I$ acquired upon measurements of the state
$\mid \psi  >$ is   proportional   to   the logarithm of the probability
$I =- \log _{2}(a\left|\matrix{\kappa }\right|^{2})$ bits,
where $a=\mid \psi _{0}\mid ^{-2}$.

The  certain  amount  $\epsilon$  of  energy  per  bit  is   expended,
transferred, stored, associated  with  the  information   encoded.
That's why it is necessary to add the  term  $\epsilon I$
to  the  standard Hamiltonian:
$$
H = H_{0} + \epsilon I = \sum^{N}_{j=1} {1\over 2m} \nabla ^{2}_{j} +
U - \epsilon \log _{2}(a\mid \psi \mid ^{2})
\eqno(2)
$$
Now it is possible to show that this constant $\epsilon$ is very close
to  our  compressibility  constant.  The  total  energy   as   the
expectation value of the Hamiltonian (2) is:
$$ { E=<\psi \mid H> = <\psi \mid H_{0} \mid \psi  >+
\epsilon <\psi \mid H_{0}\mid \psi  > = }
$$
$$
{= E_{0} \epsilon <\psi \mid \log _{2}
(a\mid \psi \mid^{2}) \mid \psi > } \eqno(3)
$$
Defining the entropy for the subsystem:
$$
S = K<\psi \left|\matrix{\ln (a}\right|\psi \mid ^{2})\mid \psi  >
\eqno(4)
$$
Where $K$ - the  Boltzmann  constant. Now  we  obtain  the  free
energy $E_{0}$, the temperature $T$ and constant $\epsilon$ :
$$
E_{0} = E - TS,\quad T=\left( {\partial E\over \partial S} \right) _{V},
\quad \epsilon =  KT \ln(2) \eqno(5)
$$
The value $\epsilon$ is very close to  the  constant  C.  If  we
suppose the value $KT$ for the average  kinetic energy of
nucleons $\approx 30 MeV$  we obtain the reasonable value of
compressibility constant.

Besides NoSE there is  another  interesting  equation,  which  was
applied many times for the prediction of new forms of  the  motion
in nuclei\cite{r12}. This  is 
well-known Korteveg de Vries Equation  (KdV)  equation.  
KdV  equation  was
applied many times in nuclear physics\cite{r12,r13,r14,r15,r16,r17}.

In the work\cite{r12} the connection was established  between  the
solution of the KdVE and  the  nuclear  potential  in  Schroedinger
equation. At the certain circumstances we can see the evolution of
the nuclear potential with time.

\section{Nonlinear nuclear vortex}
In heavy ion collisions a localized stable or long lived excitation might
develop. The existence of a fairly well-defined nuclear surface makes it
possible to discuss localized excitations in the nuclear surface region.

G.N.Fowler, S.Raha and R.M.Weiner\cite{r15} attempted  to  predict
the localized stable surface excitations in peripheral  heavy  ion
collisions like in Rossby waves.

In the work\cite{r16} A.Sandulesky and W.Greiner  obtained  a  new
coexistence model consisting  of  the  usual  shell  model  and  a
cluster model, describing a soliton moving on the nuclear surface.
They got the excellent description of experimental   spectroscopic
factors for  cluster  decays.

In the framework of semiclassical nonlinear nuclear hydrodynamics\cite{r17}
we have presented a type of vortical excitations new for nuclear physics.
In the paper \cite{kar93jaf}
we began to consider
a pure vortical motion in nuclear systems
excluding the usual approximation of  smallness of the excitation
amplitude and the additional assumptions about the shape of
the nuclear system.
We considered the two-dimensional analog of a nuclear disk
\cite{kar92}, a plane nuclear vortex, a new type of a pure vortical state
of incompressible inviscid nuclear matter.
It is a finite area of  constant vorticity in a plane
within a uniformly rotating contour.
These states can be considered as the generalization of the
elliptic Kirchhoff vortex \cite{Lamb}.
Recently we generalized this type of solution
to the similar three-dimensional picture and consider
a new possible class of
nonlinear Euler type equations on a nuclear surface\cite{lend95m}.
We have found a class of uniformly rotating solutions to nonlinear Euler
type equations on a sphere. The equations of motion describing
a localized vortex on a spherical nuclear surface -
bounded region of constant vorticity, surrounded by irrotational flux -
are derived\cite{iopvort96}.
These excitations may be linked
 with  hot spots created in peripheral heavy ion collisions\cite{hotspot}.

Let us consider the next type of nonlinear excitations in the region of
nuclear surface,
as a quadrupole deformation of the nuclear potential:
$$
U = U_{0}\hbox{ sech}^{2}((r-R)/a) = 2aU_{0}{df(2(r-R)/a)\over dr} \eqno(6)
$$
The evolution of this perturbation having  the  soliton  form
with time is given by the KdV equation\cite{r18}:
$$
U_{t} - 6UU_{x}+ \beta U_{\hbox{xxx}}= 0,\qquad \eqno(7)
$$
where the function $U$ is the potential of Schroedinger equation
$$
\beta \psi _{xx} - U\psi  = 0, \qquad \beta  = \hbar^{2} / 2m \eqno(8)
$$
On the other hand this perturbation in the  form  of  soliton
can decay on the certain  quantity  of  solitons  with  the  small
amplitude\cite{r19,r20}. If we fix the KdVE in the form
$$
U_{t}+ UU_{x}+ \beta U_{\hbox{xxx}}= 0,\eqno(9)
$$
its solution will be in the form:
$$
U(x)=U_{0}\hbox{ sch}^{2} (U_{0}/12\beta )^{1/2} X) \eqno(10)
$$
Now we apply the similarity principle. The initial condition  we
take in the form:
$$
U(x,0) = U_{0}\phi (x,l).\eqno(11)
$$
With new variables
$$
\eta = U/U_{0},\quad x = X/l, \quad \tau = U_{0}t/l,\eqno(12)
$$
we obtain the new KdVE:
$$\eta_{t} + \eta \eta_{x} + \sigma^{-2} \eta_{\hbox{xxx}} = 0,\eqno(13)$$
where $\sigma = l(U_{0}/\beta ) ^{1/2}.$

In the work\cite{r19} the critical value $\sigma = \sqrt{12}$  was  obtained.
With increasing number $\sigma$ the corresponding perturbation breaks  up
into a larger number of solitons. The number of this solitons  was
determined by Karpman\cite{r20}
$$N = 2a \sqrt{\frac{U_0}{6\beta}}. \eqno(14)$$

For the scattering of $\alpha$-particles with kinetic  energies  in
the range 4 and 12 MeV we can get for the nucleus $^{12}C$ the value $N
{\sim} 5.$  It corresponds to the number of existing  states   generated
by the surface potential well: $K^{+} = 0, 2, 4, 6, 8$. Gardner et
al.\cite{r21} showed,  that  the  amplitude  of  solitons  determined
by eigenvalues: $A = 2 E_{n} $.

The potential having the soliton shape  gives  the  following
spectrum\cite{r22}:
$$
E_{n}  = - {\hbar^{2} \over 8m \Delta}
\left( -(1-2n) + (1 + 8m\Delta U_{0} {\hbar^{-2}})^{1/2} \right). \eqno(15)
$$
For our  example $(n=0,2,4,6,8)$  we  can  get  the  following
values ($MeV$): $-15.57$, $-8.65$, $-1.75$, $5.19$, $12.11$
respectively. The smaller eigen value the lager soliton amplitude.
Knowing the eigen values one can  construct  the  total  potential
\cite{r23}.

This expression (6) provides the key to the understanding of the behaviour
of the deformation length for different targets and for different
projectiles. In a recent paper by Koster et. al.\cite{r25} the
empirical formula
$$ {(\beta_2 r_0)}_{nucleons} = { Z \over A_T}{(\beta_2 r_0)}_{protons}
 + { N \over A_T}{(\beta_2 r_0)}_{neutrons}   \eqno(16)
$$
is given to explain the different values of the deformation length
for even isotopes of Pd\cite{r26}.

In terms of the soliton approach the compatibility condition for solitons
in different media can be written\cite{r27}
$$ A \delta = A' \delta ' = const  \eqno(17)
$$
where $A$ and $A\prime$ correspond to the formula (17) and
$$ \delta = \frac{\hbar^2}{2\mu} \hspace{1cm}
  \delta ' = \frac{\hbar^2}{2\mu '}$$
and
$$  \mu=\frac{A_{p,n}\cdot A_T}{A_{p,n}+A_T} \hspace{1cm}
    \mu '=\frac{A_{p,n}\cdot A_T'}{A_{p,n}+A_T'}
$$

The product $A\delta$ should be the same for different isotopes
of the target nucleus. The resulting prediction is that the
deformation length in the case of inelastic scattering of
$\alpha$-particles on different Pd isotopes is should be largest for
the heavest one for the 2$^+$ state.

Thus we interpreted two equations NoSE and KdVE  respectively
to the  problem  of  clustering.  This  work  is  only  the  first
approximation to this problem because we used only one  dimension.
To  our  mind   the   nextstep   should   be the   application   of
Kadomtsev-Petviashvily equation to this problem.

The most promising thing is the application of the Similarity principle
to the linear alpha-chain states in nuclei\cite{r24}.

 \section{Solitons as the clusters}

What  are clusters?   Clusters  are spatially  correlated
nucleons.   What  kind of  forces are  between clusters  and
clusters and nuclei?  Japanes theoreticians showed\cite{aa1},
that there are  four typical "di-molecules" in  nature (moleculas
consisting  from  atoms,   nuclei,  nuclei  plus  hyperon,
nucleons) with the same kind of potential.   We consider the
typical potential interaction between $^4 He$ atoms.

   In the approximation  one can consider $ He$  atoms as hard
spheres with the radius $2.7 A$.  Minimum says that atoms have
a tendency  to develop into  crystall.  But the  warm energy
does not permit it.

   In the  different energy  scales the  situations is  very
similar.

   The origination of  the attractive part of  the potential
is due  to the exchange of  nucleons in the case  of nuclear
potential\cite{aa2}.

   We shall  consider the repulsive  core.  The  best understanding
can be obtained in  the micro and macro approaches.
The existence of a repulsive  core in the potential interaction
of two  complex nuclei  follows  from Pauli  principle
which forbidds  their mutual interpenetration.  Let  us give
one example~--~the interaction  of two  alpha-particles.  In
this interaction the  repulsive core is manifested  as a
result  of  which  quasimolecular state  the  nucleus  formes.
The  $\alpha + \alpha$ phase  shifts  were obtained from
the different experiments.  The phase shift $\delta _2 $ begins to increase
rapidly from an energy about $1\ MeV$  and  $\delta _4 $ from an energy about
$ 6\ MeV $.   This indicates  an extension  with a  fairly abrupt
boundary in  the exterrior  region with $ r \leq 4 fm $  (the impact
parameter $b = \lambda l$ is approximately $4\ fm$ at $E_{cm} = 1\ MeV$
 and $3\ fm$ at $E_{cm} =6\ MeV $ ). This character of  the scattering
can be well described by a phenomenological  potential with the repulsive
core.

In terms of the NoSE we can estimate the value of a hard core.
For this purpose it is necessary to rewrite the NoSE.

$$
-\frac{\hbar^2}{2m}\,\frac{d^2}{dx^2} U  + ( E - V ) U + \beta U^3 = 0 .
\eqno(18)
$$

After the factorization at the wave funtion $U$
$$
U= \psi_0 f,
$$
$$
f=0,\ at\ x=0,
$$
$$
f=1,\ at\ x \rightarrow \infty
$$
we obtain the NoSE
$$
\frac{\hbar^2}{2 m}\,\frac{d^2}{dx^2} \psi  + \alpha \psi + \beta \psi ^3 = 0.
\eqno(19)
$$
Then using such approaches $\alpha \rightarrow \infty, U = \psi_0, \psi^2_0 =
- \alpha / \beta $ we have
$$\frac{\hbar^2}{2 m \alpha}\,\frac{d^2}{dx^2} \psi + 
\frac{\beta}{\alpha} \psi ^3 = 0.
\eqno(20)
$$
In the compact form this equation can be transformed
into another equation
$$
- \xi^2 \frac{d^2}{dx^2} f -f + f^3 =0.
\eqno(21)
$$
Multiplaying this equatin on the derivative $df/dx$ we can obtain
a solution
$$
                f = th [ x/ \sqrt 2 \xi].
\eqno(22)
$$
The solution is very similar on the correlation function
of Boze gas with hard spheres:
$$
                f=0,\ for\ r < \delta,
$$
$$
                f=th[r/ \delta - 1 ].
$$
Where  $\delta$ -- is the radius of a hard sphere.
    In terms of this approach the radius of the hard core is:
$$
R_0 = \sqrt 2 \xi = \sqrt 2 \frac { \hbar^2}{ \sqrt {2m(E-V)} }
\frac { \hbar^2 } { \sqrt {m(E-V)} } \,.
\eqno(23)
$$

The microscopic estimates of the value of the hard core
were made in the work of V.N.Bragin and R.Donangelo\cite{aa15}
and in the work of A.Tohsaki-Suzuki and K.Naito\cite{aa16}~.
The last athors used RGM. In this case it is very evident
to expect the appearance of the hard core. In the first
work the athors used the Brukner theory according to whitch
the total energy of a system of interacting fermions can be
written as a functional of the local single-particle energy
density. They have computed the heavy ion potential
in the sudden approximation as
$$
V(r)= \int [ \xi  (\rho_1 + \rho_2) - \xi (\rho_1) - \xi (\rho_2) ]
 d{\tau}\,.
\eqno(24)
$$

The potential which was obtained in this approuch is highly
repulsive in the nuclear interior quite an opposite behavior
to that of the folding model calculations.
Summing up, it is necessary to say, that the momentum and
energy dependent hard or soft core appears in the
microseoptic calculations of two interacting nuclei due to
Pauli prinsiple. On this base this core can be taken in
account also in the macroscopic approaches.

Keeping in mind the structure of the potential for nucleous--nucleous
collision (the repulsive and attractive parts) we applied the inverse
field method, which will be described in the next section. Within the
inverse mean field method solitons are taken to model of elastic
$\alpha+\alpha$ collisions in a TDHF--like fasion. Attention is drawn
to common points of this approach with TDHF. The analytical formula
for the phase-shift within this approach yields a nice correspondence
to experiment\cite{aa17}.

\section{Nuclear instability and soliton theory}
An existence of solitary waves is
determined by two essential factors, namely, nonlinearity and dispersion.
Both the factors, which are responsible for the stability
of a wave, are connected in their turn
with two different types of instability.
A localized pulse will tend to spread out due to dispersion terms of
the equations of motion.
The nonlinearity which is responsible for
the formation of solitary waves, on the other hand,
automatically leads to their destruction, if it is alone.
Both instabitities may compensate each other
and lead to stable solutions (solitons).

Let us look from these points of view at multifragmentation phenomena.
The formation and breakup of a highly excited and compressed nuclear
system is the most striking process
observed in intermediate-energy heavy ion reactions. How does such a system
expand and finally disassemble when passing through a regime of dynamical
instabilities? What is the mechanism of the clustering (stable
light and intermediate mass fragment production)?
Quite a variety of models have been developed to discuss
this question (see, for instance,
the recent review\cite{morwoz93} and Proceedings\cite{hirschegg94}).
The dynamical clusterization in the presence of instabilities
is the focus of attention of the intermediate energy
heavy ions physics\cite{knoll,burchorun92}.
Ten years ago multifragmentation has been associated with the onset of
the spinodal instability\cite{si83}. This instability
is associated with the transit of a homogeneous fluid across a domain of
negative pressure, which leads to its breaking up into droplets of
denser liquid embedded in a lower density vapor. Since the spinodal
instability can occur in an infinite system, it can be called the bulk or
volume instability. On the other hand, it physically means that  pressure
depends on density, that is just a nonlinearity in terms of density
\cite{kargreece}.

Recently\cite{mtcw92}, it has been pointed out that a new kind of instability
(sheet instability)
may play an important role in multifragmentation. This new instability
can be assigned to the class of surface instabilities of
the Rayleigh-Taylor kind\cite{rayleigh}.
System escapes from the high surface energy of the
intermediate complex
by breaking up into
a number of spherical fragments with less overall surface.
At the same time, it physically means the existence of the gradient
terms of the equations of motion, i.e the dispersion\cite{kargreece}.

The spinodal instability and the Rayleigh-Taylor instability
may compensate each other and lead to stable quasi-soliton type objects.
In the next Sec. we present this physical picture
using a simple
analytical model proposed to describe the time evolution of compressed
nuclear systems.

An analysis of stability of nonlinear
dynamical systems and an analysis of nonlinear evolution of initial
complex states is a traditional goal of  Soliton Theory.
The inverse methods to integrate nonlinear evolution equations
are often more effective than a direct numerical integration.
Let us demonstrate this statement for a very simple case\cite{kar88}.

The type of systems under our consideration are uncharged slabs of
nuclear matter.
The slabs are finite in the $z$ coordinate and
infinite and homogeneous in two transverse directions.

The basic equations for the static mean-field description of the slabs
are the following
$$ \psi_{{{\bf k}_{\perp}} n}
({\bf x}) = {1 \over {\sqrt{\Omega}}} {\psi}_{n} (z)
\exp (i {\bf k}_{\perp}{\bf r}),\qquad
\epsilon_{{{\bf k}_{\perp}} n} = {\hbar^{2} k_{\perp} ^{2}
\over 2m} + e_{n}, $$
$$-{\hbar^{2} \over 2m} {d^{2} \over dz^{2}}
\psi_{n} (z) + U(z) \psi_{n} (z) = e_{n} \psi _{n} (z), $$
$$
\rho ({\bf x}) \Longrightarrow \rho (z) =
\sum _{n=1} ^{N_{0}} a_{n} \psi _{n} ^{2} (z), \eqno(25)
$$
$$
A \Longrightarrow {\cal A} = ( 6 A {\rho _{N} ^{2}} / \pi ) ^{1/3} =
\sum _{n=1} ^{N_{0}} a_{n}, \qquad
a_{n} = {2m \over \pi \hbar^{2} } ( e_{F} - e_{n}),
$$
$$
{E \over A} \Longrightarrow { \hbar^{2}
\over 2m{\cal A}} \Bigr(
 \sum _{n=1} ^{N_{0}} a _{n} \int _{-\infty} ^{\infty}
\bigr( {d \psi_{n} \over dz} \bigl) ^{2} dz
+ {\pi \over 2} \sum _{n=1} ^{N_{0}} a_{n} ^{2} \Bigl)
 + {1 \over {\cal A}} \int _{-\infty} ^{\infty}
{\cal E} [ \rho (z) ] dz,
$$
where ${\bf r} \equiv (x,y), \/ {{\bf k}_{\perp}}
\equiv (k_{x}, k_{y})$, and $\Omega$ is the transverse normalization area.
$ N_{0}$ is the number of occupied bound orbitals.

The dynamical description will be done in the framework of the inverse mean
field method. One can found the details of this approach
in\cite{hef84},\cite{hk87}.
We concentrate here only on essentials.

The evolution of a system is given by the famous hydrodynamical
Korteweg-de Vries equation (KdV) for the mean-field potential
$U(z,t)$
$$
\sum\limits_{n=1}^{N} {\partial U \over {\partial (S_{n}t)}} =
6 U {\partial U \over \partial z} - {\hbar^{2} \over 2m}
{\partial ^{3} U \over \partial z ^{3}}, \eqno(26)
$$
where $S_{n}$ are constants which are determined by the initial conditions.

General solution of KdV Eq.~(26) can be derived in principle
via direct methods numerically.
This way is to assign a functional of interaction ${\cal E}$
(as usual an effective density dependent Skyrme force), a total number of
particles (or a thickness of a slab ${\cal A}$) and to solve Hartree-Fock
equations to derive a spectrum of the single particle states
$e_{n}$ and wave functions $\psi _{n} (z,0)$, the density profile
$\rho (z,0)$ and the one-body potential $U(z,0)$ for the initial
compressed nucleus. Then, one calculate an evolution of
the one-body potential with the help of Eq.~(26).

However, there is an inverse method to solve KdV Eq.~(26).
The main advantage of this way is to reduce the solution of the nonlinear
KdV Eq.~(26) to the solution of the linear Schroedinger - type  equation
$$
-{\hbar^{2} \over 2m} {d^{2} \over dz^{2}}
\psi_{n} (z,t) + U(z,t) \psi_{n} (z,t) = e_{n} \psi _{n} (z,t),\eqno(27)
$$
and linear integral Gelfand - Levitan - Marchenko equation
to derive the function $ K ( x , y )$
$$
K (x, y) + B (x + y) + \int _{x} ^{\infty} B (y + z) K (x,z) dz =0.
\eqno(28)
$$
The kernel $B$ is determined by the reflection coefficients
$R (k) \/ ( e_{k} = \hbar^{2} k^{2} / 2m )$.
and by the $N$ bound state eigenvalues
$e_{n} = - \hbar^{2} \kappa_{n} ^{2} / 2m $.
$N $ is the total number of the bound orbitals.
$$B (z) = \sum _{n=1} ^{N} C_{n} ^{2} ( \kappa_{n} )  +
{1 \over \pi} \int _{-\infty} ^{\infty} R (k) exp \/ (ikz) dk.$$
The coefficients $C _{n}$ are uniquely specified by the boundary conditions
$$C _{n} (\kappa _{n}) = \lim_{z \rightarrow \infty} \psi_{n} (z)
\/ \exp (\kappa_{n} z),$$
and the wanted single particle potential is given by
$$U (z,t) = - {\hbar^{2} \over m} {\partial \over \partial z} K (z, z).$$
The time $t$ is included in Eqs.~(27,28) only as a parameter, so it
has been omitted in the above formulas.

The general solution, $U(z,t)$, should naturally contain both,
contributions due to the continuum of the spectrum and to its discrete
part. There is no way to obtain the general solution $U(z,t)$ in a closed
form. Eqs.~(27,28) have to be solved only numerically.

However in the case of reflectless ($R(k) = 0$)
symmetrical ($U(-z,0) = U(z,0)$) potentials
one can derive the following basic relations
$$U (z,t) = - {\hbar^{2} \over m} {\partial^{2} \over \partial z^{2}}
\ln ( \det \Vert M \Vert ) = - {2 \hbar^{2} \over m}
\sum _{n=1} ^{N} \kappa_{n} \psi _{n} ^{2} (z,t),$$
$$\psi _{n} (z,t) = \sum _{n=1} ^{N} ( M^{-1} )_{nl} \lambda _{l} (z,t),$$
$$
\lambda _{n} (z,t) = C_{n} (\kappa_{n}) \exp \/ ( - \kappa_{n} z +
2{\hbar}^{2} {\kappa}_{n} ^{3} S_{n} t / m)), \eqno(29)
$$
\eject
$$M_{nl} (z,t) = \delta_{nl} + {{\lambda_{n} (z,t) \lambda_{l} (z,t)} \over
{\kappa_{n} + \kappa_{l}}},$$
$$C_{n} (\kappa_{n}) = \Bigl( 2 \kappa_{n} {\prod _{l {\not=} n} ^{N}}
{{\kappa_{n} + \kappa_{l}} \over {\kappa_{n} - \kappa_{l}}} \Bigr) ^{1/2}.$$
So, the wave functions, potential and the density profile are completely
defined by the bound state eigenvalues.

The first step is to solve the Schroedinger Eq.~(27) for the initial
potential $U(z,0)$, which is suitable to simulate compressed nuclear
system or to simulate this state with the help of spectrum.
Then one calculates the evolution of $\rho (z,t)$ and $U(z,t)$
with the help of Eqs.~(29).


Although there is definitely some progress in the application of the
inverse methods to nuclear physics
they are not yet too popular. As illustration
of these methods, we considered a one-dimensional three-level system
in details\cite{hol94}.

A three-level system may be useful for modelling the evolution of light
nuclei, for instance, of oxygen\cite{ddg}.

for large $z$ and $t$ the time - dependent one - body potential and
the corresponding density distributions are represented by a set of
stable solitary waves. The energy spectrum of an initially compressed
system completely determines widths, velocities and the phase shifts
of the solitons.

The number of waves is equal to the number of occupied bound orbitals.
Thickness ('number' of particles) of an $n$ - wave is equal to $a_{n}$.

Reflecting terms ($R(k) \neq 0$) of $U(z,t)$ cause ripples
(oscillating waves of a small amplitude) in addition to the solitons
in the final state.

The initially compressed system expands so that for large times one
can observe separate density solitons and ripples ('emissions').
This picture is in accordance with the TDHF simulation of the time
evolution of a compressed $O^{16}$ nucleus\cite{ddg}. The disassembly shows
collective flow and clusterization.
It is important to note that the clusterization was not observed in the
absence of the self-consistent mean-field potential, i.e this confirmes
our supposition that the nonlinearity is very important for the clustering.

It is necessary to note that the present model is too primitive
in order to describe a real breakup process. However this model can be
used to illustrate an inverse mean-field method scheme,
a nonlinear principle of superposition and the idea
that nonlinearity and dispersion terms of the evolution equation
can lead to clusterization in the final channel\cite{ijmpe94}.

Certainly there are a lot of open questions relative to
the presented approach. The most crucial concerns the generalization
to 3+1 dimensional model with a finite temperature.
One possible way to do it keeping symplicity of the approach
would be to use an information theory\cite{dridgreece}.
Such investigations are in progress\cite{lend95},\cite{spec95}.

\eject
\section{Summary}
It is shown that very different effects of nuclear cluster physics
could be explained in the frame of Korteveg de Vries Equation and
cubic Nonlinear Schr\"odinger Equation.

The NoSE has been shown to be a very useful tool for the analysis
of experimental data involving the elastic interaction of light and heavy
ions. Its novel future is that it determines fairly accurately the
compressibility modulus of nuclear matter. The problem of dynamical
instability and clustering are considered from the points
of view of the soliton conception and using KdV type equation for
a mean-field potential. Nonlinear nuclear vortex is discussed..
Due to the soliton conception it is understandable the behavior of
the quadrupole deformation parameters for different isotopes of
one nucleus.

\end{document}